\documentstyle[preprint,aps]{revtex}
\begin{document}
\title{Imaging Electron Wave Functions of Quantized Energy Levels in Carbon
Nanotubes}
\author{Liesbeth C. Venema, Jeroen W. G. Wild\"{o}er, Jorg W. Janssen, Sander J.
Tans, Hinne L. J. Temminck Tuinstra, Leo P. Kouwenhoven, Cees Dekker\thanks{%
To whom correspondence should be addressed. email: dekker@qt.tn.tudelft.nl.}}
\address{}
\maketitle

\begin{abstract}
Carbon nanotubes provide a unique system to study one-dimensional
quantization phenomena. Scanning tunneling microscopy is used to observe the
electronic wave functions that correspond to quantized energy levels in
short metallic carbon nanotubes. Discrete electron waves are apparent from
periodic oscillations in the differential conductance as a function of the
position along the tube axis, with a period that differs from that of the
atomic lattice. Wave functions can be observed for several electron states
at adjacent discrete energies. The measured wavelengths are in good
agreement with the calculated Fermi wavelength for armchair nanotubes.

\vspace{0.6in} \hspace{-0.25in}L. C. Venema, J. W. G. Wild\"{o}er, J. W.
Janssen, S. J. Tans, H. L. J. Temminck Tuinstra, L. P. Kouwenhoven, C.
Dekker, \ 

\hspace{-0.25in}Department of Applied Physics and DIMES, Delft University of
Technology, Lorentzweg 1, 2628 CJ\ Delft, The Netherlands.
\end{abstract}

\newpage 

\noindent Carbon nanotubes are molecular wires that exhibit fascinating
electronic properties \cite{Dressel}. Electrons in these cylindrical
fullerenes are confined in the radial and circumferential directions and can
only propagate in the direction of the tube axis. Nanotubes are therefore
interesting systems for studying the quantum behavior of electrons in one
dimension (1D). Limiting the length of a carbon nanotube leads to a
'particle-in-a-box' quantization of the energy levels. Such discrete energy
levels have been observed in transport experiments on individual nanotubes
and ropes \cite{Bock,Tans}. The electron wave functions corresponding to
these discrete states can in principle be imaged by Scanning Tunneling
Microscopy (STM). The well-known STM\ work on quantum corrals demonstrated
the possibility to directly image wave patterns in the local density of
states of a 2D metal surface \cite{Crommie}. Here, we apply this technique
to map out the wave functions of single molecular orbitals in short metallic
carbon nanotubes. Electronic wave functions are apparent from periodic
oscillations in the low-bias differential conductance along the tube axis.
To our knowledge, this is the first time that wave functions of discrete
electron states have been imaged in a molecular wire.

Previous STM spectroscopy studies were done at a large ($\sim 2$ eV) energy
scale to investigate the bandstructure of nanotubes\cite{nature,Odom}. These
experiments confirmed the prediction \cite{Hamda} that carbon nanotubes can
be semiconducting or metallic depending on the tube diameter and the chiral
angle between the tube axis and hexagon rows in the atomic lattice. In this
paper we focus on the low-energy ($\sim 0.1$ eV) features of short metallic
nanotubes which exhibit quantum size effects. Single-wall nanotubes with a
diameter of about $1.4$ nm were deposited on Au(111) substrates \cite
{nature,thess}. On most tubes we were able to obtain STM images with atomic
resolution \cite{nature} which allows to determine the chiral angle and
diameter of the tubes\cite{note1}. The nanotube of Fig. 1B is identified as
an `armchair' tube from the good fit between the observed hexagon structure
and the overlay of the graphene lattice. Armchair tubes have a nonchiral
structure because the hexagon rows are parallel to the tube axis. This type
of tubes has metallic properties \cite{Hamda}. Current-voltage $I(V)$
characteristics measured up to $\pm 0.5$ V on the armchair tube of Fig. 1B
indeed demonstrate the simple linear behavior expected for a metallic tube.
Such $I(V)$ measurements are done by keeping the STM tip stationary above
the nanotube, switching off the feedback, and recording the current as a
function of the voltage applied to the sample. In all our experiments the
STM is operated at 4.2 K \cite{jeroen}.

In the experiments reported here, the armchair tube of Fig. 1B was shortened
to a length of about $30$ nm. This was achieved by locally cutting the tube
by applying a voltage pulse of $+5$ V to the STM tip at a position on the
tube located 30 nm from its end \cite{apl}. STM spectroscopy was then
carried out near the middle of the short tube. $I(V)$ curves on the
shortened nanotube show a step-like behavior (Fig. 2A), which we ascribe to
quantum size effects. Steps in $I(V)$ correspond to quantized energy levels
entering the bias window upon increasing the voltage. Current steps at a
voltage $V$ thus correspond to discrete electron states at energy $%
E=E_{F}+\alpha eV$, with $E_{F}$ the Fermi energy, $e$ the electron charge,
and $\alpha \simeq 1$ \cite{notealfa}. The experimentally observed width of
the current plateaus between the steps ranges from $0.05$ V to $0.09$ V. The
plateau width is determined by the total energy to add an electron to the
tube. This addition energy consists of a combination of finite-size level
splitting and the Coulomb charging energy that is due to the small
capacitance of the tube. A simple estimate for the energy level splitting
for a tube of length $L=30$ nm is given by $\Delta E=hv_{F}/2L=0.06$ eV,
where $v_{F}=8.1\cdot 10^{5}$ m/s is the Fermi velocity and $h$ is Plancks
constant. The capacitance of a nanotube lying on a metallic substrate can be
approximated by the formula for a metallic wire parallel to a conducting
plane, $C=2\pi \epsilon _{0}L/\ln [(d+(d^{2}-R^{2})^{1/2})/R]$\cite
{capaciteit}, where $\epsilon _{0}=8.85\cdot 10^{-12}$ F/m, $d$ is the
distance from the wire axis to the plane and $R=0.65$ nm is the wire radius.
Estimating $d\approx 0.9$ nm gives $C\approx 2.0$ aF, which yields a
charging energy $E_{c}=e^{2}/C=0.08$ eV. Both numbers are in the same range
as the observed plateau width. Since the charging energy and level splitting
are of about equal magnitude, an irregular step spacing in the $I(V)$ curve
is expected\cite{leo}. For this report, the main point is that each step
corresponds to a discrete energy level entering the bias window.

The central result of our experiments is that the tunneling conductance
measured for such discrete states is found to oscillate along the length
direction of the nanotube with a period different from the atomic lattice
constant. $I(V)$ spectroscopy curves were obtained at different locations on
top of the nanotube along a line parallel to the tube axis in
constant-current mode. At every point, spaced $23$ pm apart, the feedback
was switched off to take an $I(V)$ curve, starting at the bias voltage used
for feedback in the constant-current mode. Figure 2A shows several $I(V)$
curves obtained in this way at different positions. The current displays a
clear variation between maximum (dashed curves) and minimum values (solid
curves) for negative bias voltage. Peaks in the differential conductance $%
dI/dV$ (Fig. 2B) appear at the voltage positions of current steps in the $%
I(V)$ curves$.$ The height of the $dI/dV$ peaks varies periodically with
position $x$ along the tube axis, as shown in Fig. 2C. The period of these
oscillations in the differential conductance is about $0.4$ nm, which
clearly differs from the lattice constant of $0.25$ nm. The periodic
variation of $dI/dV$ versus $x$ can - as discussed below in detail - be
attributed to the electronic wave functions in the nanotube.

The wave functions of several adjacent energy levels can be displayed
simultaneously by plotting the differential conductance $dI/dV$ as a
function of the voltage and the position $x$ along the tube (Fig. 3A). Wave
patterns can be observed for 4 different energy levels appearing at bias
voltages $0.11$ V$,$ $0.04$ V$,$ $0.00$ V and $-0.05$ V\cite{charging}. At
each level a horizontal row of about 7 maxima is resolved in $dI/dV$ as a
function of position $x$ along the tube (see Fig. 3B for the 1D spatial
profile of the wave functions belonging to these states). The experimental
quantity $dI/dV$ is a measure for the squared amplitude of the quantized
electron wave function $|\psi (E,x)|^{2}$\cite{bloch}. The curves in Fig. 3B
are fitted with a function of the form $dI/dV=A$sin$^{2}(2\pi x/\lambda
+\phi )+B$, which represents a simple trial function for $|\psi (E,x)|^{2}$.
The separation of about $0.4$ nm between peaks in $dI/dV$ corresponds to
half the wavelength $\lambda $, because $dI/dV$ measures the square of the
wave function. The wavelengths obtained from the fitting procedure vary from 
$0.66$ to $0.76$ nm (Fig. 3B). Other measurements on the same tube
reproduced values for $\lambda $ in the range of $0.65-0.8$ nm. From
repeated spectroscopy measurements such as Fig. 3A on the same tube, we
estimate the error in the wavelength to be about $0.02$ nm. Note that the $%
dI/dV$ maxima in Fig. 3A occur at different positions $x$ for the different
horizontal rows. This excludes many experimental artifacts such as for
example oscillations in the STM and provides compelling evidence for the
interpretation in terms of standing electron waves. Typically only a small
number ($\sim 4$) of discrete levels were observed around zero bias. At
larger bias voltages (beyond the images shown here), peaks in $dI(V)/dV$
could no longer be discerned clearly. At these voltages the broadening of
energy states apparently exceeds their separation. Similar electron waves
with a wavelength of about $0.7$ nm were also observed in a number of other
shortened metallic nanotubes. On shortened semiconducting nanotubes the
level splitting could not be resolved and attempts to measure electron waves
were unsuccessful. A small energy level splitting is indeed expected for
semiconducting tubes since here the Fermi energy is located at the top of a
band\cite{nature}.

Figure 3C shows the topographic height profile from the constant-current
measurement at $+0.3$ V, which clearly has a different periodicity from that
observed in $dI/dV$ (Fig. 3B). The period of $0.25$ nm is in agreement with
the atomic lattice constant $a_{0}=0.246$ nm for an armchair nanotube.
Apparently we image the atomic corrugation at high bias voltage.
Simultaneously, $I(V)$ spectroscopy curves are measured at every point (Fig.
3B and C), starting at the set-point used for feedback ($100$ pA and $+0.3$
V). As a result from maintaining feedback at this voltage, the lattice
periodicity is largely compensated because the STM tip follows the atomic
corrugation, which makes it possible to resolve the quantized electron waves
in Fig. 3B and C\cite{bloch}.

Discrete levels are probed at energies near the Fermi energy $E_{F}$, and
therefore the wavelength of the electron waves is close to the Fermi
wavelength $\lambda _{F}$. Electronic bandstructure calculations \cite
{Dressel,Hamda} for armchair tubes yield two bands near $E_{F}$ with a
linear energy dispersion $E(k)=E_{F}\pm %TCIMACRO{\UNICODE[m]{0x127}}
%BeginExpansion
\rlap{\protect\rule[1.1ex]{.325em}{.1ex}}h%
%EndExpansion
v_{F}(k-k_{F})$, where $%TCIMACRO{\UNICODE[m]{0x127}}
%BeginExpansion
\rlap{\protect\rule[1.1ex]{.325em}{.1ex}}h%
%EndExpansion
=h/2\pi $, $k=2\pi /\lambda $ is the wave vector, and $k_{F}=2\pi /\lambda
_{F}$ is the Fermi wave vector. In undoped nanotubes, the two bands cross at
the Fermi energy where $k=k_{F}=2\pi /3a_{0}$. This yields $\lambda
_{F}=3a_{0}=0.74$ nm, independent of the length of the tube. For nanotubes
on Au(111) however, $E_{F}$ is shifted away from the crossing point to lower
energy by $\delta E=0.3$ eV. This is due to charge transfer as a result of a
difference in workfunction with the underlying substrate \cite{nature}. This
shifts $k_{F}$ to $k_{F}\pm \delta k$ with $\delta k=\delta E/
%TCIMACRO{\UNICODE[m]{0x127}}
%BeginExpansion
\rlap{\protect\rule[1.1ex]{.325em}{.1ex}}h%
%EndExpansion
v_{F},$ and $\lambda _{F}$ thus becomes $\frac{2\pi }{k_{F}\pm \delta k}=0.69
$ nm $(+)$ or $0.79$ nm $(-)$. The experimentally observed wavelengths (Fig.
3B) correspond well to the theoretical values, confirming the predicted band
structure with two linear bands crossing near $E_{F}$. This result provides
quantitative evidence for our interpretation of the oscillations in $dI/dV$
in terms of wave functions of discrete electron states.

A short metallic nanotube resembles the textbook model for a particle in a
1D box. For a discrete energy state with quantum number $n$, the
corresponding wavelength $\lambda _{n}=2L/n$. The observed wavelength is
much smaller than the tube length, in accordance with the fact that the
number of electrons within one nanotube band is large ($n\sim 10^{2}$). The
wavelength will therefore vary only slightly ($\lambda _{n}/n\sim 0.01$ nm)
for adjacent discrete energy levels in one band. 

The measurements reported here are technically challenging because they
require a large series of reproducible $I(V)$ curves. Occasionally, we were
able to resolve some of the spatial structure in the wave function at a
length scale smaller than the Fermi wavelength, as shown in Fig. 4A. In this
scan the peak spacing is nonequidistant, leading to an apparent `pairing' of
peaks. This feature indicates that the wave function does not conform to a
simple sinusoidal form. Recent calculations by Rubio et al. \cite{Rubio}
indicate a nontrivial spatial variation of the nodes in the wave function of
discrete electron states in the direction perpendicular to the tube axis
(compare Fig. 4B). Line profiles can either show pairing or an equidistant
peak spacing depending on the exact position of the line scan. The
observation of pairing confirms that the relevant period in the line scans
is the distance between next-nearest-neighbor peaks.

Our experiments demonstrate that individual wave functions corresponding to
the quantized energy levels in a short metallic nanotube can be resolved
because of the large energy level splitting. The technique for recording the
wave periodicity at different energy states provides a tool for further
exploration of the dispersion relation in nanotubes. Future work will
include similar experiments on nanotubes with various chiral angles. The
methodology presented in this paper also opens up the possibility of
obtaining full 2D spatial maps of the electron wave functions in carbon
nanotubes.

\newpage

Figure captions

\mathstrut

Fig. 1. STM topographic images of individual single-wall carbon nanotubes. (%
{\bf A}) Example of a nanotube which has been shortened by applying a
voltage pulse to the STM tip above the tube \cite{apl}. ({\bf B}) Atomically
resolved image of an armchair nanotube. The arrow denotes the direction of
the tube axis. This nanotube can be identified as armchair-type because the
hexagon rows run parallel to the direction of the tube axis (cf. overlay of
the graphene lattice). The tube diameter is 1.3 nm. This image has been
taken before the tube was shortened to 30 nm. Feedback parameters are $V=0.1$
V, $I=$ 20 pA. Images were taken in constant-current mode.

\mathstrut

Fig. 2. STM spectroscopy measurements on a $30$ nm long armchair tube. ({\bf %
A}) Current-voltage $I(V)$ characteristics on the tube shown in Fig. 1B,
taken at positions about $0.18$ nm apart [data points 1 to 4 in (C)] on a
straight line along the tube axis. Current steps correspond to discrete
energy states entering the bias window. ({\bf B}) Differential conductance $%
dI/dV$ versus $V$, as calculated from the $I(V)$ curves. Peaks appear at the
voltage positions of current steps in the $I(V)$ curves. ({\bf C})
Differential conductance $dI/dV$ as a function of position along the tube.
Data was taken at a bias voltage of $-0.08$ V. Data points 1 to 4 indicate
the positions at which the four $I(V)$ curves plotted in (A) and $dI/dV$
curves in (B) were obtained.

\mathstrut

Fig. 3. Spectroscopy and topography line scans along the nanotube which show
electron wave functions of discrete electron states as well as the atomic
lattice. ({\bf A}) Differential conductance $dI/dV$ (in color scale) against
the bias voltage ($y$-axis) and the position on a straight line along the
tube ($x$-axis). This plot results from about 100 $I(V)$ curves taken at
positions about $23$ pm apart along the tube axis. Electron wave functions
of 4 different energy levels are observed as periodic variations in $dI/dV$
along the tube at voltages of $0.11$ V$,$ $0.04$ V$,$ $0.00$ V and $-0.05$
V. A horizontal row of about 7 $dI/dV$ maxima is observed at each energy
level. Note that the exact voltages at which the peaks in $dI/dV$ appear in
Fig. 2 and Fig. 3A are different\cite{charging}. ({\bf B}) $dI/dV$ profiles
at the 4 resolved energy levels. Fits of the function $dI/dV=A$sin$^{2}(2\pi
x/\lambda +\phi )+B$ are plotted as dashed curves. The resulting wavelengths 
$\lambda $ are given above the curves on the right. Curves are vertically
offset for clarity. ({\bf C}) Topographic height profile $z(x)$ of the
nanotube. STM topographic imaging and spectroscopy was performed
simultaneously by scanning the tip along the tube and recording both the tip
height (with feedback on) and $I(V)$ spectroscopy curves (feedback off).
Feedback parameters are $V=0.3$ V and $I=100$ pA.

\mathstrut

Fig. 4. Pairing of conductance peaks. ({\bf A}) Spectroscopy line scan where
pairing of $dI/dV$ maxima can be observed. Neighboring peaks are
nonequidistant, indicating a nonsinusoidal wave function. The distance
between next-nearest-neighbor peaks is approximately $0.75$ nm which agrees
with the Fermi wavelength. ({\bf B})\ Schematic of a possible arrangement of
lobes of the wave function of a single molecular orbital. In a line scan
along the blue line, peaks in $dI/dV$ will be equidistant, whereas pairing
will occur if a line scan is carried out along the red line.

\mathstrut


\begin{references}
\bibitem[9.]{note1}  Exact identification of the exact lattice indices $(n,m)
$ (see ref.\cite{Dressel}) of chiral nanotubes is quite difficult since both
the nanotube diameter and the chiral angle between hexagon rows and the tube
axis have to be measured with high accuracy. The $(n,m)$ indices are crucial
for chiral tubes because a minor change in one of these determines whether
the tube is a metal or a semiconductor. For armchair tubes, however, the
situation is easier. The non-chiral structure with hexagon rows running
parallel to the tube axis can be easily observed from the atomically
resolved images, and a precise determination of the diameter is not
essential because armchair tubes of all diameters are metallic.
\bibitem[12.]{notealfa}  The applied voltage is in principle divided into a
part that drops over the tunnel gap and a part that drops between the
nanotube and the substrate. The ratio $\alpha $ between these voltages is
determined by the capacitance ratio. Because the capacitance between
nanotube and substrate is much larger than that between the nanotube and the
STM tip, the voltage will drop almost entirely over the tunnel gap, and
accordingly $\alpha $ has a value close to 1.
\bibitem[16.]{bloch}  The total wavefunction is in fact defined by the
atomic lattice potential modulated with a standing wave profile resulting
from the confinement in the length direction. Because the STM tip follows
the atomic corrugation by scanning in constant-current mode at a high bias
voltage, the lattice periodicity is largely compensated so that the standing
waves can be resolved in the spectroscopy measurements for several discrete
states at low bias.
\end{references}
\end{document}